\documentclass[12pt]{article}

\usepackage{graphicx}

\newcommand{\beq}{\begin{equation}}
\newcommand{\eeq}{\end{equation}}
\newcommand{\bea}{\begin{eqnarray}}
\newcommand{\eea}{\end{eqnarray}}

\newcommand{\gsim}{\lower.7ex\hbox{$
\;\stackrel{\textstyle>}{\sim}\;$}}
\newcommand{\lsim}{\lower.7ex\hbox{$
\;\stackrel{\textstyle<}{\sim}\;$}}

\renewcommand{\Im}{{\rm Im}\,}

\def\ot{{\bf T}}
\def\cp{{\bf CP}}
\def\cpt{{\bf CPT}}

\begin{document}

\thispagestyle{empty}
\vspace*{-10mm}

\begin{flushright}
UND-HEP-05-BIG\hspace*{.08em}02\\
DPNU-05-07\\
LAL 05-48\\
hep-ph/0506037 \\
%\today \\
%{\tiny TAUnewIkeV22.tex}
\end{flushright}
\vspace*{4mm}

%%%%%%%%%%%%
\centerline{\large\bf A `Known' \cp~Asymmetry in $\tau$ Decays} 
%%%%%%%%%%%%%
\vskip 0.3cm \centerline{I. I. Bigi$^a$ and A.I. Sanda$^b$} 

\vskip 0.3cm 
\centerline{$^a$Dept. of Physics, University of Notre Dame du Lac, Notre Dame, IN 46556, U.S.A.}
\centerline{$^b$Dept. of Physics, Nagoya University, Nagoya, Japan} 
\centerline{e-mail: ibigi@nd.edu, sanda@eken.phys.nagoya-u.ac.jp}
\vskip 0.5cm
\centerline {\em Dedicated to our late colleague and friend P. Kabir}
\vskip 0.5cm

\centerline{\bf Abstract}
We point out that dynamics known from the observed \cp~violation in 
$K_L\to \pi^{\mp}l^{\pm}\nu$ coupled with \cpt~invariance 
generate a \cp~asymmetry of $3.3\cdot 10^{-3}$ in 
$\tau^{\pm} \to \nu K_S\pi^{\pm}\to \nu [\pi^+\pi^-]_K\pi^{\pm}$. 
The equality of the $\tau^+$ and $\tau^-$ lifetimes required by \cpt~symmetry is 
restored in an intriguing way as the combined effect of long-lived 
$K\to \pi^+\pi^-$ as well as contributions from $K_L-K_S$ interference.  
While little new can be learnt from this \cp~asymmetry, the latter has to be 
accounted for, since \cp~asymmetries in $\tau \to \nu K\pi$ channels are prime candidates for revealing 
the intervention of New Physics. This `known' \cp~asymmetry provides a very useful 
calibration point in such searches. It also provides a test of \cpt~symmetry (as well as the 
$\Delta Q = \Delta S$ selection rule).

%\newpage
%%%%%%%
\section{Introduction}
%%%%%%%%%

With the CKM prediction of large \cp~asymmetries in $B$ decays 
\cite{CS,BS80} like $B_d \to \psi K_S$ confirmed \cite{BELLEBABAR}, the main task of $B$ factories of all stripes is to look for `New Physics', i.e.  dynamics beyond the Standard Model (SM). This goal is being pursued by studying $B$ decays of ever 
greater rarity. We want to stress that $\tau$ decays likewise deserve extensive efforts for three reasons 
at least: 
\begin{itemize}
\item 
No \cp~violation has been observed yet in leptodynamics. Finding it there would represent a 
qualitative step forward. 
\item 
There are intriguing scenarios, where baryogenesis is driven by leptogenesis as primary effect 
\cite{LEPTO}. \cp~violation is then required in leptodynamics. This is the main justification for undertaking 
Herculean efforts to find \cp~violation in neutrino oscillations. Searching for \cp~asymmetries in 
$\tau$ decays provides another of the few meaningful avenues towards that goal. 
\item 
Like for $B$ mesons, studies of $\tau$ decays very likely provide different and presumably complementary perspectives onto the anticipated New Physics connected with the electroweak phase transition.

\end{itemize} 
An optimal environment for studying $\tau$ decays is provided by $e^+e^- \to \tau ^+\tau ^-$. It 
offers a high rate relative to  all other final states in a `clean' and well-understood environment 
that allows searching for SM forbidden modes like $\tau ^{\pm} \to l^{\pm}\gamma$, $\mu^{\pm}l^+l^-$ 
with $l =e,\mu$. Maybe even more importantly is another unique opportunity such $\tau$ factories 
offer, whether they are of the $\tau$-charm or $B$ factory of Giga-Z variety: they enable 
searches for novel \cp~asymmetries. Since the $\tau$ pair is produced with its spins aligned, one can 
use the decay of one $\tau$ to `tag' the spin of the other $\tau$ and thus probe for spin dependent 
\cp~asymmetries {\em without} needing polarized beams. 

In this short note we want to point out that contrary to a wide spread perception known dynamics 
generate \cp~asymmetries in $\tau$ decays: the well-measured \cp~asymmetry in 
$K_L \to \pi^{\mp}l^{\pm}\nu$ produces a difference in 
$\Gamma (\tau^+ \to \overline \nu K_L\pi^+)$ vs. $\Gamma (\tau^- \to \nu K_L\pi^-)$, where 
$K_L$ is defined as the neutral kaon decaying on a time scale $\sim {\cal O}(\Gamma_L^{-1})$, 
and -- assuming \cpt~symmetry -- the same asymmetry also in 
$\Gamma (\tau^+ \to \overline \nu K_S\pi^+)$ vs. $\Gamma (\tau^- \to \nu K_S\pi^-)$ with the  
$K_S$ defined as the neutral kaon decaying on the much shorter time scale 
$\sim {\cal O}(\Gamma_S^{-1})$. We explain how the apparent conflict with 
\cpt~invariance enforcing equal $\tau ^+$ and $\tau ^-$ lifetimes is resolved. Such \cp~asymmetries, 
which of course are absent in  
$\Gamma (\tau^+ \to \overline \nu K^+\pi^0)$ vs. $\Gamma (\tau^- \to \nu K^-\pi^0)$, have  
to be taken into account; it also provides a powerful calibration, when searching for 
manifestations of New Physics through \cp~studies. 

%%%%%%%%%%
\section{SM \cp~violation in $\tau$ decays}
%%%%%%%%%%

The SM predicts for the transition amplitudes 
\beq
T(\tau^-\to \overline K^0 \pi^- \nu )=T(\tau^+\to K^0 \pi^+ \overline \nu) \; , 
\label{K0K0BAR}
\eeq 
since there is no weak phase and the strong phase has to be the same. 
Yet the observed kaons are the mass and not the flavour eigenstates, i.e. 
$K_S$ and $K_L$, rather than $K^0$ and $\overline K^0$. Ignoring \cp~violation 
in $\Delta S\neq 0$ dynamics, one has $\langle K_L|K_S\rangle =0$, and the 
$K_L$ and $K_S$ are unambiguously distinguished by their decay modes in addition to their 
vastly different lifetimes: $K_S \to 2 \pi$, $K_L\not\to 2\pi$, $K_L\to 3\pi$. Then one has 
\bea 
\Gamma (\tau ^- \to \nu K_S \pi^-) &=& \Gamma (\tau ^- \to \nu K_L \pi^-) = 
\frac{1}{2}\Gamma (\tau ^- \to \nu \overline K^0 \pi^-) \\
\Gamma (\tau ^+ \to \overline \nu K_S \pi^+) &=& \Gamma (\tau ^+\to \overline \nu K_L \pi^+) = 
\frac{1}{2}\Gamma (\tau ^+ \to \overline \nu  K^0 \pi^+)
\eea
and thus no \cp~asymmetry due to Eq.(\ref{K0K0BAR}). 

The situation becomes considerably more complex and intriguing, once \cp~violation in 
$\Delta S=2$ transitions is included. (We can safely ignore {\em direct} \cp~violation for our purposes here.) Imposing \cpt~invariance we can write
\bea
|K_S\rangle&=&p|K^0\rangle +q|\overline K^0\rangle 
\nonumber
\\ 
|K_L\rangle &=&p|K^0\rangle -q|\overline K^0\rangle 
\label{KLKSDEFCPT}
\eea
with $|p|^2 + |q|^2 = 1$. We then have 
\beq 
\langle K_L|K_S\rangle = |p|^2 - |q|^2 \simeq 2 {\rm Re}\epsilon_K \simeq 
(3.27 \pm 0.12)\times 10^{-3} \neq 0
\eeq 
as deduced from 
\beq 
\frac{\Gamma(K_L\to \pi^- l^+\nu)-\Gamma(K_L\to \pi^+ l^- \overline \nu)}
{\Gamma(K_L\to \pi^- l^+\nu)+\Gamma(K_L\to \pi^+ l^- \overline \nu)} = 
|p|^2 - |q|^2 
\eeq
The mass eigenstates thus are no longer orthogonal to each other and both 
$K_S \to 2\pi$ and $K_L\to 2\pi$ can occur. I.e., the $2\pi$ final state by itself no longer 
distinguishes strictly between $K_S$ and $K_L$. Yet the difference in lifetimes still 
provides a discriminator: Considering the decay rate evolution for $\tau \to \nu \pi [\pi^+\pi^-]_K$ as a function of $t_K$, the (proper) time of the kaon decay, one has for  
{\em short} decay times -- $t_K \sim {\cal O}(\Gamma_S^{-1})$ -- we have for all practical purposes 
only $K_S \to 2\pi$ decays and find 
\beq 
\frac{\Gamma(\tau^+\to [\pi^+\pi^-]_{"K_S"} \pi^+ \overline \nu)-
\Gamma(\tau^-\to [\pi^+\pi^-]_{"K_S"} \pi^- \nu)}
{\Gamma(\tau^+\to [\pi^+\pi^-]_{"K_S"}\pi^+ \overline \nu)+
\Gamma(\tau^-\to [\pi^+\pi^-]_{"K_S"} \pi^- \nu)}=
|p|^2-|q|^2\simeq (3.27 \pm 0.12)\times 10^{-3} \; . 
\label{KS}
\eeq
For {\em long} decay times -- $t_K \sim {\cal O}(\Gamma_L^{-1})$ -- we have 
again for all practical purposes 
only $K_L \to 2\pi$ and find 
\beq 
\frac{\Gamma(\tau^+\to [\pi^+\pi^-]_{"K_L"} \pi^+ \overline \nu)-
\Gamma(\tau^-\to [\pi^+\pi^-]_{"K_L"} \pi^- \nu)}
{\Gamma(\tau^+\to [\pi^+\pi^-]_{"K_L"}\pi^+ \overline \nu)+
\Gamma(\tau^-\to [\pi^+\pi^-]_{"K_L"} \pi^- \nu)} = |p|^2 - |q|^2
\label{KL}
\eeq
Strictly speaking it does not even matter which forces generate $|q| \neq |p|$, whether it is due to 
to SM dynamics or not, as long as $\tau$ decays {\em themselves} are described by the SM.  

Measuring the asymmetry of Eq.(\ref{KL}) seems hardly feasible, since the $K_L$ acts basically like a 
second neutrino, yet it raises an intriguing question: With the asymmetries in 
Eqs.(\ref{KS},\ref{KL}) having the same sign, how is the equality of the $\tau^+$ 
and $\tau^-$ lifetimes restored as required by \cpt~invariance, 
which we have explicitly invoked? 

To answer this question, let us recall how we might actually measure
the asymmetries of Eqs.(\ref{KS},\ref{KL}). These asymmetries are obtained by studying the elapsed time between the $\tau$ decay and the time 
at which $"\pi\pi"$ is formed. The first asymmetry is obtained by looking at events with short time difference, where as the second asymmetry is obtained by looking decays with large time difference. \cpt~constraint applies only to the total decay rate where we include events at all times of decay.

Because $\langle K_L|K_S\rangle \neq 0$, the 
decay rate evolution for $\tau \to \nu \pi [f]_K$,where $f$ is an arbitrary final state, now contains three terms: in addition to the two 
contributions listed above with a time dependance $\propto e^{-\Gamma_St_K}$ and 
$\propto e^{-\Gamma_Lt_K}$, respectively, we have an interference term 
$e^{-\frac{1}{2}(\Gamma_S + \Gamma_L)t_K}$ most relevant for intermediate 
times $\Gamma_S^{-1} \ll t_K \ll \Gamma_L^{-1}$. 
\footnote{It was this interference term in $K^0 (t) \to \pi^+\pi^-$ and 
$\overline K^0(t) \to \pi^+\pi^-$, which established originally that the Fitch-Cronin observation 
of $K_L \to \pi^+\pi^-$ could not be reconciled with \cp~symmetry by suggesting that they had actually 
observed $K_L \to \pi^+\pi^-U$ with $U$ denoting a hitherto unknown neutral particle with odd 
\cp~parity that had escaped detection.} 
Note that
because of the interference term, observing only $\pi\pi$ final state does not allow us to understand the 
\cpt~constraint. Measuring all three terms for all $f$ and integrating over all $t_K$ -- possible in principle, though 
maybe not  
in practice -- recovers the full information on the production of $\overline K^0$ and 
$K^0$ with the relation of Eq.(\ref{K0K0BAR}).

To be more explicit: one has to track the full decay rate evolution into a general state $f$ when the initial state 
was a $K^0$ -- $\Gamma (K^0(t_K)\to f)$ -- versus a 
$\overline K^0$ -- $\Gamma (\overline K^0(t_K)\to \overline f)$. 
The most general expression reads 
\bea 
\nonumber 
\Gamma (K^0((t_K))\to f) =&& \frac{1}{2|p|^2} \left[|T(K_S \to f)|^2  e^{-\Gamma _St_K}
+ |T(K_L \to f)|^2 e^{-\Gamma _Lt_K}  
\right. \\
&+& \left.2e^{-\frac{1}{2}(\Gamma _S + \Gamma _L)t_K} {\rm Re}(e^{i\Delta M_Kt_K}T(K_S \to f)T(K_L \to f)^*) \right] 
\label{GENEXP1}
\\ 
\nonumber 
 \Gamma (\overline K^0(t_K)\to \overline f) =&& \frac{1}{2|q|^2} \left[|T(K_S \to \overline f)|^2  e^{-\Gamma _St_K}
+ |T(K_L \to \overline f)|^2 e^{-\Gamma _Lt_K}  
\right. \\
&-& \left.2e^{-\frac{1}{2}(\Gamma _S + \Gamma _L)t_K} {\rm Re}(e^{i\Delta M_Kt_K}T(K_S \to \overline f)T(K_L \to \overline f)^*) \right] 
\label{GENEXP2}
\eea
For short times of decay the first term dominates, which describes $K_S$ decays, and 
Eq.(\ref{KS}) applies; for very long times the second term does producing the same \cp~asymmetry 
as stated in Eq.(\ref{KL}). Yet for the intermediate range in times of decay the third term reflecting 
$K_S-K_L$ interference becomes important.

By rewriting the {\em interference} term in terms of $K^0$ and $\overline K^0$,
integrating over $t_K$ and finally summing over all possible final state
$f$ and $\overline f$, we have
\bea
\nonumber 
&&\frac{1}{|p|^2}\sum_f\int dt_Ke^{-\frac{1}{2}(\Gamma _S + \Gamma _L)t_K}
%{\rm Re} 
[e^{i\Delta M_Kt_K} T(K_S \to  f)T(K_L \to  f)^*)]\\
\nonumber 
&&=%{\rm Re} \left[ 
\frac{1}{\frac{\Gamma_L+\Gamma_S}{2}-i\Delta M} [2(|p|^2-|q|^2)\Gamma_{11}+i4 {\rm Im}(qp^*\Gamma_{12})]\\% \right] \\
&&=%{\rm Re}\left[ 
2(|p|^2-|q|^2)+\frac{2i}{\frac{\Gamma_L+\Gamma_S}{2}-i\Delta M}[2\Delta M~{\rm Re}\epsilon -\Delta\Gamma~{\rm Im}\epsilon],% \right] , 
\label{int1}
\eea
where we have used relations valid for this problem to first order in the \cp~violating
parameters: $\Gamma_{11}=\frac{\Gamma_L+\Gamma_S}{2}$, 
$p=\frac{1}{\sqrt{2}}(1+\epsilon)$, $q=\frac{1}{\sqrt{2}}(1-\epsilon)$, 
 $\Delta \Gamma=2\Gamma_{12}$. Finally using   
 ${\rm arg}~\epsilon={\rm arc tan}\left(\frac{2\Delta M}{\Delta\Gamma}\right)$ we find that the 
 square bracket in the last line of Eq.(\ref{int1}) vanishes; i.e. 
\beq
\frac{1}{|p|^2}\sum_f\int dt_Ke^{-\frac{1}{2}(\Gamma _S + \Gamma _L)t_K}
[e^{i\Delta M_Kt_K} T(K_S \to  f)T(K_L \to  f)^*)]=2(|p|^2-|q|^2)
\label{int2}
\eeq
Using Eq.(\ref{int2}) it is simple to show that the interference term indeed restores the constraints from \cpt~symmetry:
\beq 
\sum_f\int dt_K \Gamma(\tau^+\to [f]_{"K^0(t_K)"} \pi^+ \overline \nu)
=\sum_{\overline f}\int dt_K \Gamma(\tau^-\to [\overline f]_{"\overline K^0(t_K)"}  \pi^- \nu).
\eeq
While this is as it must be, it is still instructive to see how it comes about. 

In talking about the time of decay $t_K$ we were referring to the proper time of the neutral kaon decay. 
In a real experiment one has two times of decay, namely that of the $\tau$ lepton and of the kaon. 
The explicit formulae have been given for the even more involved case of 
$D^0 \to K_SX$ allowing even for $D^0 - \overline D^0$ oscillations to take place \cite{AZI98}. 
Yet for the experimentally most accessible case involving $K_S$ decays at short values of $t_K$ 
there is no practical need for the full machinery. 

%%%%%%%%%%%%%%%%%%%%%%%%%%%%
\section{\cpt~violation}
%%%%%%%%%%%%

While we are not suggesting that \cpt~violation is likely to surface in $\tau$ decays, 
it is not inappropriate to address this issue. It 
has been searched for extensively in semileptonic K decays; thus it is convenient to 
employ the notation used there. Without imposing \cpt~invariance Eq.(\ref{KLKSDEFCPT}) 
is generalized to 
\bea
|K_S\rangle&=&p_S|K^0\rangle +q_S|\overline K^0\rangle 
\nonumber
\\ 
|K_L\rangle &=&p_L|K^0\rangle -q_L|\overline K^0\rangle 
\label{KLKSDEFNOCPT}
\eea
with \cite{bs}
\bea 
p_S=N_S\cos\theta/2,&~~~~&q_S=N_Se^{i\phi}\sin\theta/2
\nonumber\\
p_L=N_L\sin\theta/2,&~~~~&q_L=N_Le^{i\phi}\cos\theta/2
\eea
where $\phi$ and $\theta$ are both {\em complex} numbers constrained by the discrete symmetries as 
follows: 
\bea 
\cpt~ \; \; {\rm or} \; \; \cp~{\rm invariance} &\Longrightarrow& \; \; 
{\rm cos}\theta = 0 
\\ 
\cp~ \; \; {\rm or} \; \; \ot~{\rm invariance} &\Longrightarrow& \; \; 
\phi = 0 
\eea
The normalization constants $N_S$ and $N_L$ are given by: 
\beq 
N_S=  \frac{1}{\sqrt{\left|{\rm cos}\frac{\theta}{2}  \right|^2 +
\left|e^{i\phi}{\rm sin}\frac{\theta}{2}  \right|^2  }}  \; , \; 
N_L = \frac{1}{\sqrt{\left|{\rm sin}\frac{\theta}{2}  \right|^2 +
\left|e^{i\phi}{\rm cos}\frac{\theta}{2}  \right|^2  }}  
\eeq

If \cpt~symmetry is violated $\cos\theta\ne 0$ and
$\Im \phi\ne 0$. 
\bea
\frac{\Gamma(\tau^+\to "K_S" \pi^+ \overline \nu)-\Gamma(\tau^-\to "K_S" \pi^- \nu)}
{\Gamma(\tau^+\to "K_S" \pi^+ \overline \nu)+\Gamma(\tau^-\to "K_S" \pi^- \nu)}&=&\Im\phi+{\rm Re}\cos\theta
\nonumber
\\
\frac{\Gamma(K_L\to \pi^- l^+\nu)-\Gamma(K_L\to \pi^+ l^- \overline \nu)}
{\Gamma(K_L\to \pi^- l^+\nu)+\Gamma(K_L\to \pi^+ l^- \overline \nu)}&=&\Im\phi-{\rm Re}\cos\theta
\label{CPTTEST}
\eea
where as before $"K_S"$ is understood as the {\em short}-time component in $K\to \pi^+\pi^-$; 
we have also assumed the $\Delta S=\Delta Q$ selection rule for these 
decay amplitudes.
We look forward to new information on ${\rm Re}\cos\theta$ from $\tau$ decays.

To be consistently heretical, one could also entertain the idea of the $\Delta S=\Delta Q$ rule being 
violated and possibly by different amounts in $K$ and $\tau$ decays. The relevant expressions 
are quite straightforward and can be derived using
similar techniques described, for example, in Ref.\cite{bs}. We will not write them down here, since we feel there is even less space for heresy in $\Delta S=1$ than in $\Delta S=2$ dynamics. 
%%%%%%%%%

%%%%%%%%%%%%
\section{Decays of other particles}
%%%%%%%%%%

The power of $K_{L,S}$ to discriminate matter against antimatter affects the decays of other particles as well. In $B_d/\overline B_d \to \psi K_S$ its effect is covered up by the huge \cp~violation in 
$\Delta B\neq 0$ dynamics. It is of relevance when SM forces generate no or only small \cp~violation like in $D^{\pm} \to K_S\pi^{\pm}$ \cite{YAMA} or in $D^{\pm} \to l^{\pm}\nu K_S$. 
%%%%%%%%%
\section{Conclusion}
%%%%%%%%%%

As stated before, the asymmetries expressed in Eqs.(\ref{KL},\ref{KS}) have to be there, since 
they are caused by a well established effect, namely that the $K_{L,S}$ are ever so slightly, yet definitely sensitive to the matter-antimatter distinction. In principle it does not even matter, 
whether the SM can reproduce the observed size of $\epsilon_K$. In that sense observing this asymmetry does not teach us anything we do not already know. 

This, however, is an incomplete evaluation of the situation. For \cp~asymmetries in the channels 
$\tau \to \nu K + {\rm pions}$ are natural portals for the emergence of New Physics. For the final state 
is sufficiently complex to allow for \cp~odd observables also in distributions beyond fully integrated widths; secondly they should be particularly sensitive to non-SM charged Higgs exchanges 
\cite{KUHN}. Obviously it is then essential to note that known physics produces a reliably predicted asymmetry in the full width, but not in final state distributions for some channels. 

There are some more subtle points as well: it is actually most useful experimentally when not all modes 
are predicted to exhibit a null effect within the SM. For if one does not observe the effect predicted 
by Eqs.(\ref{KL},\ref{KS}), then there has to be New Physics, which (partially) cancels the effect from 
known physics -- or one does not understand one's experiment with the required accuracy. The small 
expected \cp~asymmetry discussed above thus provides a most helpful calibration. 

Finally it behooves us to allow for the admittedly exotic possibility of \cpt~invariance being violated 
in general and in the dynamics of the $K^0 - \bar K^0$ complex in particular. Eq.(\ref{CPTTEST}) 
provides a novel test for it. 

\vspace{0.5cm}

{\bf Acknowledgments:} One of us (I.B.) gratefully acknowledges the hospitality extended to him by the Laboratoire de l'Acc\' el\' erateur Lin\' enaire, while this work was done. We thank our colleagues J. Rosner and Y. Grossman for pointing our attention to the problem of how equal lifetimes arise and 
Ya. Azimov for providing us with references to his work. 
This work was supported by the NSF under grant PHY03-55098 and by JSPS under grant C-17540248.

%%%%%%%%%%%

\end{document}